\documentclass[superscriptaddress,showpacs,amssymb,amsmath,amsfonts,aps,
prl,twocolumn
]{revtex4}
\usepackage{graphicx}
\usepackage{dcolumn}
\usepackage{epsfig}
\begin{document}

\title{ Comment on the narrow structure reported by Amaryan {\it et al.}  
}


\newcommand*{\ANL}{Argonne National Laboratory, Argonne, Illinois 60439}
\newcommand*{\ANLindex}{1}
\affiliation{\ANL}
\newcommand*{\ASU}{Arizona State University, Tempe, Arizona 85287-1504}
\newcommand*{\ASUindex}{2}
\affiliation{\ASU}
\newcommand*{\UCLA}{University of California at Los Angeles, Los Angeles, California  90095-1547}
\newcommand*{\UCLAindex}{3}
\affiliation{\UCLA}
\newcommand*{\CSUDH}{California State University, Dominguez Hills, Carson, CA 90747}
\newcommand*{\CSUDHindex}{4}
\affiliation{\CSUDH}
\newcommand*{\CANISIUS}{Canisius College, Buffalo, NY}
\newcommand*{\CANISIUSindex}{5}
\affiliation{\CANISIUS}
\newcommand*{\CMU}{Carnegie Mellon University, Pittsburgh, Pennsylvania 15213}
\newcommand*{\CMUindex}{6}
\affiliation{\CMU}
\newcommand*{\CUA}{Catholic University of America, Washington, D.C. 20064}
\newcommand*{\CUAindex}{7}
\affiliation{\CUA}
\newcommand*{\SACLAY}{CEA, Centre de Saclay, Irfu/Service de Physique Nucl\'eaire, 91191 Gif-sur-Yvette, France}
\newcommand*{\SACLAYindex}{8}
\affiliation{\SACLAY}
\newcommand*{\CNU}{Christopher Newport University, Newport News, Virginia 23606}
\newcommand*{\CNUindex}{9}
\affiliation{\CNU}
\newcommand*{\UCONN}{University of Connecticut, Storrs, Connecticut 06269}
\newcommand*{\UCONNindex}{10}
\affiliation{\UCONN}
\newcommand*{\FFU}{Fairfield University, Fairfield, CT 06824}
\newcommand*{\FFUindex}{11}
\affiliation{\FFU}
\newcommand*{\FIU}{Florida International University, Miami, Florida 33199}
\newcommand*{\FIUindex}{12}
\affiliation{\FIU}
\newcommand*{\FSU}{Florida State University, Tallahassee, Florida 32306}
\newcommand*{\FSUindex}{13}
\affiliation{\FSU}
\newcommand*{\Genova}{Universit$\grave{a}$ di Genova, 16146 Genova, Italy}
\newcommand*{\Genovaindex}{14}
\affiliation{\Genova}
\newcommand*{\GWUI}{The George Washington University, Washington, DC 20052}
\newcommand*{\GWUIindex}{15}
\affiliation{\GWUI}
\newcommand*{\ISU}{Idaho State University, Pocatello, Idaho 83209}
\newcommand*{\ISUindex}{16}
\affiliation{\ISU}
\newcommand*{\INFNFR}{INFN, Laboratori Nazionali di Frascati, 00044 Frascati, Italy}
\newcommand*{\INFNFRindex}{17}
\affiliation{\INFNFR}
\newcommand*{\INFNGE}{INFN, Sezione di Genova, 16146 Genova, Italy}
\newcommand*{\INFNGEindex}{18}
\affiliation{\INFNGE}
\newcommand*{\ORSAY}{Institut de Physique Nucl\'eaire ORSAY, Orsay, France}
\newcommand*{\ORSAYindex}{19}
\affiliation{\ORSAY}
\newcommand*{\ITEP}{Institute of Theoretical and Experimental Physics, Moscow, 117259, Russia}
\newcommand*{\ITEPindex}{20}
\affiliation{\ITEP}
\newcommand*{\JMU}{James Madison University, Harrisonburg, Virginia 22807}
\newcommand*{\JMUindex}{21}
\affiliation{\JMU}
\newcommand*{\KNU}{Kyungpook National University, Daegu 702-701, Republic of Korea}
\newcommand*{\KNUindex}{22}
\affiliation{\KNU}
\newcommand*{\UNH}{University of New Hampshire, Durham, New Hampshire 03824-3568}
\newcommand*{\UNHindex}{23}
\affiliation{\UNH}
\newcommand*{\OHIOU}{Ohio University, Athens, Ohio  45701}
\newcommand*{\OHIOUindex}{24}
\affiliation{\OHIOU}
\newcommand*{\NIU}{Northern Illinois University, Dekalb, IL 60115}
\newcommand*{\NIUindex}{25}
\affiliation{\NIU}
\newcommand*{\RPI}{Rensselaer Polytechnic Institute, Troy, New York 12180-3590}
\newcommand*{\RPIindex}{26}
\affiliation{\RPI}
\newcommand*{\URICH}{University of Richmond, Richmond, Virginia 23173}
\newcommand*{\URICHindex}{27}
\affiliation{\URICH}
\newcommand*{\MSU}{Skobeltsyn Nuclear Physics Institute at Moscow State University, 119899 Moscow, Russia}
\newcommand*{\MSUindex}{28}
\affiliation{\MSU}
\newcommand*{\SCAROLINA}{University of South Carolina, Columbia, South Carolina 29208}
\newcommand*{\SCAROLINAindex}{29}
\affiliation{\SCAROLINA}
\newcommand*{\JLAB}{Thomas Jefferson National Accelerator Facility, Newport News, Virginia 23606}
\newcommand*{\JLABindex}{30}
\affiliation{\JLAB}
\newcommand*{\UNIONC}{Union College, Schenectady, NY 12308}
\newcommand*{\UNIONCindex}{31}
\affiliation{\UNIONC}
\newcommand*{\UTFSM}{Universidad T\'{e}cnica Federico Santa Mar\'{i}a, Casilla 110-V Valpara\'{i}so, Chile}
\newcommand*{\UTFSMindex}{32}
\affiliation{\UTFSM}
\newcommand*{\GLASGOW}{University of Glasgow, Glasgow G12 8QQ, United Kingdom}
\newcommand*{\GLASGOWindex}{33}
\affiliation{\GLASGOW}
\newcommand*{\WJC}{Washington \& Jefferson College, Washington, PA 15301}
\newcommand*{\WJCindex}{34}
\affiliation{\WJC}
\newcommand*{\WM}{College of William and Mary, Williamsburg, Virginia 23187-8795}
\newcommand*{\WMindex}{35}
\affiliation{\WM}

\newcommand*{\NOWIU}{Indiana University, Bloomington, IN 47405}
 %%%%%%%%%%%%%%% END OF Latex Macros for institute addresses  %%%%%%%%%%%%%%%%%%%%%%%%% 

\author {M.~Anghinolfi} 
\affiliation{\INFNGE}
\author {J.~Ball} 
\affiliation{\SACLAY}
\author {N.A.~Baltzell} 
\affiliation{\ANL}
\affiliation{\SCAROLINA}
\author {M.~Battaglieri} 
\affiliation{\INFNGE}
\author {I.~Bedlinskiy} 
\affiliation{\ITEP}
\author {M.~Bellis} 
\affiliation{\NIU}
\affiliation{\CMU}
\author {A.S.~Biselli} 
\affiliation{\FFU}
\author {C.~Bookwalter} 
\affiliation{\FSU}
\author {S.~Boiarinov} 
\affiliation{\JLAB}
\affiliation{\ITEP}
\author {P.~Bosted} 
\affiliation{\JLAB}
\author {V.D.~Burkert} 
\affiliation{\JLAB}
\author {D.S.~Carman} 
\affiliation{\JLAB}
\author {A.~Celentano} 
\affiliation{\INFNGE}
\author {S. ~Chandavar} 
\affiliation{\OHIOU}
\author {P.L.~Cole} 
\affiliation{\ISU}
\affiliation{\JLAB}
\author {V.~Crede} 
\affiliation{\FSU}
\author {R.~De~Vita} 
\affiliation{\INFNGE}
\author {E.~De~Sanctis} 
\affiliation{\INFNFR}
\author {B.~Dey} 
\affiliation{\CMU}
\author {R.~Dickson} 
\affiliation{\CMU}
\author {D.~Doughty} 
\affiliation{\CNU}
\affiliation{\JLAB}
\author {M.~Dugger} 
\affiliation{\ASU}
\author {R.~Dupre} 
\affiliation{\ANL}
\author {H.~Egiyan} 
\affiliation{\JLAB}
\affiliation{\WM}
\author {A.~El~Alaoui} 
\affiliation{\ANL}
\author {L.~El~Fassi} 
\affiliation{\ANL}
\author {L.~Elouadrhiri} 
\affiliation{\JLAB}
\author {P.~Eugenio} 
\affiliation{\FSU}
\author {G.~Fedotov} 
\affiliation{\SCAROLINA}
\author {M.Y.~Gabrielyan} 
\affiliation{\FIU}
\author {M.~Garcon} 
\affiliation{\SACLAY}
\author {G.P.~Gilfoyle} 
\affiliation{\URICH}
\author {K.L.~Giovanetti} 
\affiliation{\JMU}
\author {F.X.~Girod} 
\affiliation{\JLAB}
\author {J.T.~Goetz} 
\affiliation{\UCLA}
\author {E.~Golovatch} 
\affiliation{\MSU}
\author {M.~Guidal} 
\affiliation{\ORSAY}
\author {L.~Guo} 
\affiliation{\FIU}
\affiliation{\JLAB}
\author {K.~Hafidi} 
\affiliation{\ANL}
\author {H.~Hakobyan} 
\affiliation{\UTFSM}
\author {D.~Heddle} 
\affiliation{\CNU}
\affiliation{\JLAB}
\author {K.~Hicks} 
\affiliation{\OHIOU}
\author {M.~Holtrop} 
\affiliation{\UNH}
\author {D.G.~Ireland} 
\affiliation{\GLASGOW}
\author {B.S.~Ishkhanov} 
\affiliation{\MSU}
\author {E.L.~Isupov} 
\affiliation{\MSU}
\author {H.S.~Jo} 
\affiliation{\ORSAY}
\author {K.~Joo} 
\affiliation{\UCONN}
\affiliation{\JLAB}
\author {P.~Khetarpal} 
\affiliation{\FIU}
\author {A.~Kim} 
\affiliation{\KNU}
\author {W.~Kim} 
\affiliation{\KNU}
\author {V.~Kubarovsky} 
\affiliation{\JLAB}
\author {S.V.~Kuleshov} 
\affiliation{\UTFSM}
\affiliation{\ITEP}
\author {H.Y.~Lu} 
\affiliation{\CMU}
\author {I.J.D.~MacGregor} 
\affiliation{\GLASGOW}
\author {N.~Markov} 
\affiliation{\UCONN}
\author {M.E.~McCracken} 
\affiliation{\WJC}
\affiliation{\CMU}
\author {B.~McKinnon} 
\affiliation{\GLASGOW}
\author {M.D.~Mestayer} 
\affiliation{\JLAB}
\author {C.A.~Meyer} 
\affiliation{\CMU}
\author {M.~Mirazita} 
\affiliation{\INFNFR}
\author {V.~Mokeev} 
\affiliation{\JLAB}
\affiliation{\MSU}
\author {K.~Moriya} 
\altaffiliation[Current address:]{\NOWIU}
\affiliation{\CMU}
\author {B.~Morrison} 
\affiliation{\ASU}
\author {A.~Ni} 
\affiliation{\KNU}
\author {S.~Niccolai} 
\affiliation{\ORSAY}
\author {G.~Niculescu} 
\affiliation{\JMU}
\affiliation{\OHIOU}
\author {I.~Niculescu} 
\affiliation{\JMU}
\affiliation{\JLAB}
\affiliation{\GWUI}
\author {M.~Osipenko} 
\affiliation{\INFNGE}
\author {A.I.~Ostrovidov} 
\affiliation{\FSU}
\author {K.~Park} 
\affiliation{\JLAB}
\affiliation{\KNU}
\author {S.~Park} 
\affiliation{\FSU}
\author {S. ~Anefalos~Pereira} 
\affiliation{\INFNFR}
\author {S.~Pisano} 
\affiliation{\INFNFR}
\author {O.~Pogorelko} 
\affiliation{\ITEP}
\author {S.~Pozdniakov} 
\affiliation{\ITEP}
\author {J.W.~Price} 
\affiliation{\CSUDH}
\author {G.~Ricco} 
\affiliation{\Genova}
\author {M.~Ripani} 
\affiliation{\INFNGE}
\author {B.G.~Ritchie} 
\affiliation{\ASU}
\author {P.~Rossi} 
\affiliation{\INFNFR}
\author {D.~Schott} 
\affiliation{\FIU}
\author {R.A.~Schumacher} 
\affiliation{\CMU}
\author {E.~Seder} 
\affiliation{\UCONN}
\author {Y.G.~Sharabian} 
\affiliation{\JLAB}
\author {E.S.~Smith} 
\affiliation{\JLAB}
\author {D.I.~Sober} 
\affiliation{\CUA}
\author {S.S.~Stepanyan} 
\affiliation{\KNU}
\author {P.~Stoler} 
\affiliation{\RPI}
\author {W. ~Tang} 
\affiliation{\OHIOU}
\author {M.~Ungaro} 
\affiliation{\JLAB}
\affiliation{\RPI}
\affiliation{\UCONN}
\author {B~.Vernarsky} 
\affiliation{\CMU}
\author {M.F.~Vineyard} 
\affiliation{\UNIONC}
\affiliation{\URICH}
\author {D.P.~Weygand} 
\affiliation{\JLAB}
\author {M.H.~Wood} 
\affiliation{\CANISIUS}
\affiliation{\SCAROLINA}
\author {N.~Zachariou} 
\affiliation{\GWUI}
\author {B.~Zhao} 
\affiliation{\WM}

\collaboration{The CLAS Collaboration}
\noaffiliation
\date{\today}

\maketitle

In Ref. \cite{moskov}, the authors claim to observe a narrow structure 
in the mass spectrum constructed from the $(pK_L)$ system using data 
from the CLAS detector. The interpretation of this narrow structure 
given in Ref. \cite{moskov} is: 
"It may be due to the photoproduction of the $\Theta^+$ pentaquark 
or some unknown $\Sigma^*$ resonance." 
They go on to say that "it is unlikely for the observed structure 
to be due to a $\Sigma^*$ resonance".

This analysis 
was reviewed by the CLAS Collaboration, following the established 
procedures for all CLAS papers, and did not receive approval. The 
purpose of this note is to explain the reasons why that analysis 
was not approved for publication.

An extensive review of the analysis in Ref. \cite{moskov} was carried 
out by two separate committees of the Hadron Spectroscopy Physics 
Working Group in the CLAS Collaboration.  In both cases, the committees 
came to the same conclusion: the physics claims of Ref. [1] 
could not be supported.  The reasons for this conclusion are 
many-fold, but a primary concern is the lack of 
justification for the kinematic cuts used in that analysis. 

The review committees reported that 
the narrow structure appears 
only within a specific range of values of the kinematic cuts.
Here, the details are important (what cuts were varied and 
by how much) but this would require more space to document 
than a simple comment letter will allow.  We give only one example 
below, but note that the CLAS committees conducted an
extensive review of the sensitivity of the narrow structure 
to what they considered reasonable variations of the cuts
\cite{elton}.

As an example, the cut on the $t_\Theta$ variable 
(defined in Ref. \cite{moskov}) was restricted to a small region 
of the total phase space ($-t_\Theta < 0.45$ GeV$^2$).  
Without this cut, the narrow structure 
is not statistically significant.  
By examining Fig. 8 of Ref. \cite{moskov}, one can see that 
the structure is not really visible in the top spectrum (Fig. 8a), 
and only appears in Fig. 8c.
When the cut value is increased by 20\% ($-t_\Theta<0.55$) as shown 
by Fig. 8b, or decreased by 10\% ($t_\Theta<0.4$) as shown by 
Fig. 8d, then the purported structure at a mass of 1.54 GeV is 
consistent in size with other fluctuations in those spectra.

While the authors of Ref. \cite{moskov} make an argument about 
why the $t_\Theta$ cut was necessary, the CLAS Collaboration 
was not convinced.  For example, it is possible that an interference 
between the narrow structure and the background is dependent on 
the $t_\Theta$ variable, but this assumption is difficult to prove. 
The analysis of Ref. \cite{moskov} did not provide 
any evidence of interference phases. 

It is not uncommon to use kinematic cuts to reduce background 
and hence improve the signal-to-background ratio for known particles, 
but other studies \cite{klein} have shown that one must be careful to 
apply kinematic cuts which can create spurious fluctuations. 
We could argue whether the kinematic cuts used in 
Ref. \cite{moskov} are justified, but the fact remains that the CLAS 
Collaboration as a whole was not convinced that the narrow structure 
of Ref. \cite{moskov} corresponds to a real physical entity. 

At the request of the lead  author of Ref. \cite{moskov}, 
presentations were made at a CLAS Collaboration meeting
by both the authors and the review committee, followed by discussions 
and a vote on whether to publish that result as a collaboration paper.  
The outcome of the vote was to not publish this analysis.

In the end, the validity of the narrow structure claimed by 
Ref. \cite{moskov} will be determined by future experiments.  If it is  
physical resonance, as suggested by Ref. \cite{moskov}, 
then it should be reproducible. 
The evidence presented in Ref. \cite{moskov} was not 
sufficient to convince the CLAS Collaboration 
of the physics conclusions of that analysis.

\end{document}